\documentclass[twocolumn,floatfix,preprintnumbers,groupedaddress,nofootnoteinbib]{revtex4-1}
\pdfoutput=1

\usepackage{amsmath}
\usepackage{amssymb}
\usepackage{graphicx}
\usepackage{float}
\usepackage{braket}
\usepackage{hyperref}
\usepackage{pbox}
\usepackage{tabularx}

\usepackage[all]{hypcap}

\usepackage{feynmp}
\def\beq{\begin{equation}}
\def\eeq{\end{equation}}
\def\beeq{\begin{eqnarray}}
\def\eeeq{\end{eqnarray}}
\def\beg{\begin{gather}}
\def\eeg{\end{gather}}

\def\gp2{g^{\prime 2}}

\usepackage{color}

\begin{document}   
\title{Topology of the Electroweak Vacua} 

\author{Ben Gripaios} 
\email{gripaios@hep.phy.cam.ac.uk}
\affiliation{Cavendish Laboratory, J.J. Thomson Ave, Cambridge, CB3
  0HE, UK}

\author{Oscar Randal-Williams} 
\email{o.randal-williams@dpmms.cam.ac.uk}
\affiliation{DPMMS, Centre for Mathematical Sciences, Wilberforce Road, Cambridge, CB3 0WB,
UK}

\begin{abstract} 
In the Standard Model, the electroweak symmetry is broken by a complex, $SU(2)$-doublet
Higgs field and the vacuum manifold $SU(2)\times U(1)/U(1)$ has the
topology of a 3-sphere. We
remark that there exist theoretical alternatives
that are locally isomorphic, but
 in which the
vacuum manifold is homeomorphic to an arbitrary non-trivial
principal $U(1)$-bundle over a 2-sphere. These alternatives have non-trivial
fundamental group and thus feature topologically-stable
electroweak strings. An alternative based on the manifold
$\mathbb{R}P^3$ (with fundamental group
$\mathbb{Z}/2$) allows custodial protection of gauge boson masses and
their couplings to fermions, but, in common with all alternatives to $S^3$, has
a problem with fermion masses.
\end{abstract}   
\maketitle 
\section{Introduction \label{sec:intro}}
Decades of experiment have confirmed that the weak nuclear force and
the electromagnetic force are described by a gauge theory in which a
group 
  locally isomorphic to $SU(2)\times U(1)$ is non-linearly
realised in the vacuum, with only the electromagnetic subgroup $U(1) < SU(2)\times U(1)$
being linearly realised. Thus, the electroweak (EW) vacuum is
degenerate and the vacua are described by a homogeneous space
$SU(2)\times U(1)/U(1)$. 

The starting point for this Letter is the observation that there are
many ways to include $U(1)$ in $SU(2)\times U(1)$; different ways
lead to homogeneous spaces that can be topologically
inequivalent. In the Standard Model (SM), the vacuum manifold 
arises due to a non-vanishing vacuum expection value (VEV) of
the Higgs field, carrying the doublet representation of $SU(2)$, and is
homeomorphic to the 3-sphere, $S^3$. As is well-known, this is rather
boring from a physicist's point of view, since the vanishing of the homotopy groups $\pi_1 (S^3)$ (respectively
$\pi_2(S^3)$) implies the absence of  topologically-stable strings (respectively
monopoles). Here, we investigate different inclusions of $U(1)$, which lead to vacuum
manifolds with fundamental group given by an arbitrary cyclic group;
alternatives to the SM based on such inclusions thus feature
topologically-stable strings, with potentially interesting
consequences, {\em a priori}, for
astrophysics, cosmology, and particle physics. 

Given the recent discovery \cite{Aad:2012tfa} of a particle whose properties correspond
rather closely to that of the Higgs boson, consideration of
alternatives to the SM requires a willing suspension of disbelief on
the part of the reader, and indeed we shall see that none of the
topologically-distinct alternatives can be made consistent with
data. Nevertheless, we feel that the very existence of such 
theoretical alternatives to the status quo is a noteworthy curiosity in its own
right.

The outline is as follows. In \S \ref{sec:21}, we discuss the topology
of the vacuum manifold $SU(2)\times U(1)/U(1)$ and in \S
\ref{sec:nlsm} we show that an general effective field theory based on a
vacuum manifold with non-trivial topology can be consistent with
data, apart from a problem with fermion masses. In \S \ref{sec:lsm}
and \S \ref{sec:ch}, we present explicit examples with non-trivial
topology based on linear sigma models and composite Higgs models.
\section{Topology of $SU(2)\times U(1)/U(1)$ \label{sec:21}}
We begin our discussion by assuming that the EW gauge group really is
$G = SU(2)\times U(1)$, deferring discussion of groups locally
isomorphic thereto
until the end. We write elements of $G$ as $(U,z)$, where $U$ is a $2\times 2$ unitary matrix
with unit determinant and $z$ is a unit complex number. For $p,q \in \mathbb{Z}$ there is a homomorphism $\phi_{p,q} : U(1) \to G$ given by $\phi_{p,q}(z) = (\mathrm{diag}(z^q,z^{-q}),z^p)$, and if $(p,q)$ are coprime then $\phi_{p,q}$ is injective, in which case we write $H_{p,q} \leq G$ for its image. (Any injective homomorphism $\phi : U(1) \to G$ is conjugate to some $\phi_{p,q}$, as its projection to the $SU(2)$-factor may be conjugated to land in the standard maximal torus.)

Our first goal is to investigate the topology of the homogeneous spaces
$G/H_{p,q}$. An immediate result is that $G/H_{p,q}$ cannot be
homeomorphic for different $p$, because a loop wound once around
$H_{p,q}$ is wound $p$ times around the $U(1)$ factor of $G$. This implies, using the
long exact sequence of homotopy groups $\pi_1 (H_{p,q}) \cong \mathbb
Z \rightarrow \pi_1 (G) \cong \mathbb Z \rightarrow \pi_1(G/H_{p,q})
\rightarrow \pi_0 (H_{p,q}) \cong 0$ of the fibre bundle 
$H_{p,q} \hookrightarrow G \rightarrow G/H_{p,q}$, that $\pi_1
(G/H_{p,q}) \cong \mathbb{Z}/p$. Moreover, we see that 
topologically-stable string configurations occur when $p \neq 1$ \cite{Vilenkin:2000jqa}.

To investigate the topology further, let $K_{p,q} = H_{p,q} \cap (SU(2) \times \{1\}) \leq SU(2)$ and consider the function $\pi :A \mapsto (A,1) H_{p,q} : SU(2) \to G/H_{p,q}$. This is a composition of smooth maps $SU(2) \hookrightarrow G \to G/H_{p,q}$ and so smooth. The differential at the identity $D\pi : \mathfrak{su}(2) \to \mathfrak{g}/\mathfrak{h}_{p,q}$ is an isomorphism, and by homogeneity it follows that $\pi$ is a submersion, and hence a local diffeomorphism. Furthermore the right $K_{p,q}$-action on $SU(2)$ acts freely and transitively on the fibres of $\pi$, exhibiting it as a principal $K_{p,q}$-bundle, and hence giving a diffeomorphism $SU(2)/K_{p,q} \cong G/H_{p,q}$.

Now $K_{p,q} = \{\mathrm{diag}(e^{2\pi i qk/p},e^{-2\pi i qk/p}) \,:\, k \in \mathbb{Z}\}$ is the same subgroup of $SU(2)$ as $K_{p,1}$, because $(p,q)$ are coprime, and as $K_{-p,1}$: thus we shall suppose $p > 0$. It follows that $G/H_{p,q}$ is diffeomorphic to $SU(2)/K_{p,1}$, which is further diffeomorphic, as we now show, to a lens space \cite{Tietze}. These spaces are of
great historical importance in mathematics, providing the first
examples of manifolds whose homeomorphism type is determined by neither
their fundamental group and homology \cite{Alexander}, nor even
their homotopy type \cite{Threlfall}. The
lens space $L(n,m)$ is defined for $(n,m)$ coprime as the quotient of the unit sphere,
$S^3 \subset \mathbb{C}^2$ by the free $\mathbb{Z}/n$-action generated
by $(z_1,z_2) \mapsto (e^{2\pi i /n} z_1, e^{2\pi i m/n} z_2)$. Identifying $SU(2)$ with the unit sphere $S^3 \subset \mathbb{C}^2$, $SU(2)/K_{p,1}$ is thus identified with the lens space $L(p,1)$.

The lens spaces $L(p,1)$ are precisely those 3-manifolds that arise as principal $U(1)$-bundles over the 2-sphere (except for $S^2 \times U(1)$). Indeed, the clutching construction shows that such bundles are in bijection with $\pi_1(U(1)) = \mathbb{Z}$, and this bijection may be given by assigning to a principal $U(1)$-bundle over the 2-sphere its Euler number. 
Writing $U(1) = \{\mathrm{diag}(e^{i \theta},e^{-i \theta}) \,:\, \theta \in [0,2\pi)\} \leq SU(2)$, the Hopf bundle
$h_1 : SU(2) \to SU(2)/U(1) = S^2$ is the principal $U(1)$-bundle with Euler number 1. As $K_{p,1} \leq U(1)$ the map $h_1$ is the composition
$$SU(2) \longrightarrow SU(2)/K_{p,1} \overset{h_p}\longrightarrow SU(2)/U(1) = S^2$$
of a $p$-fold covering map and a principal $(U(1)/K_{p,1} \cong U(1))$-bundle $h_p$, whose Euler number is therefore $p$ and whose total space is $G/H_{p,q} \cong SU(2)/K_{p,1} \cong L(p,1)$.

From this perspective, we may use standard results to read off
the algebraic topological invariants of $G/H_{p,q}$: the homotopy groups are given by $\pi_1 =
\mathbb{Z}/p, \pi_{i>1} = \pi_{i>1} (S^3)$ (so $\pi_2 = 0, \pi_3 =
\mathbb{Z}, \pi_4 =\mathbb{Z}/2,$ \&c.); the integral cohomology is
given by $H^0 = \mathbb{Z}, H^1 = 0, H^2 = \mathbb{Z}/p, H^3 =
\mathbb{Z}$. Most interesting among these, for physicists, is $\pi_1 =
\mathbb{Z}/p$.

How do these results relate to the SM?
In that case, we postulate the
existence of a Higgs field, that is a matter field $\phi$ whose
potential is such that it
acquires an non-vanishing VEV. It carries
the doublet irreducible representation of $SU(2)$ and its charge $q
\in \mathbb{Z}$ under $U(1)$ is non-vanishing, but otherwise arbitrary.  
The $G$-action is then $G:\phi \mapsto Uz^q\phi$. Without loss of generality, we
may write the Higgs VEV as $\langle \phi \rangle = (0,
v)^T$, such that the unbroken subgroup is $H_{1,q} = \{ (\mathrm{diag}
(z^{q},z^{-q}), z)\}$. The discussion above then shows that,
as expected, the SM EW vacuum manifold is homeomorphic to $S^3$ and does not
feature topologically-stable strings.
\section{Non-linear sigma model \label{sec:nlsm}}
Having established the existence of homogeneous spaces with
non-trivial topology, we now construct physical theories based upon them,
beginning with non-linear sigma models (NLSMs). These represent the most
general low energy-effective field theories consistent with the
non-linearly realised symmetry $G$ \cite{Coleman:1969sm} and we would
like to examine whether such
theories can also be consistent with experimental data for any $p,q$. 

At the local level at least, this is almost a triviality. Indeed, even in the ungauged theory, 
physics which depends only on local
properties of the vacuum manifold can only depend on $p$ and $q$ through
their quotient. But even the quotient is unphysical (locally) in the
gauged theory, because
differing values of $p$ and $q$ can be absorbed by redefinitions of the
gauge coupling constants.

As an explicit example, consider quark masses. The necessary and sufficient
condition for writing these in the NLSM is that one can form
$H_{p,q}$-invariants out of the the quark fields $Q, U^c,$ and
$D^c$. Writing the corresponding $U(1)$ hypercharges as $y_Q, y_U,$
and $y_D$, the $SU(2)\times U(1)$ action on the $SU(2)$-doublet $Q$ restricts
under $H_{p,q}$ to $Q \mapsto \mathrm{diag} (z^{q+py_Q},
z^{-q+py_Q})Q$, while the action on the $SU(2)$-singlets $U^c$ and $D^c$
restricts to  $U^c,D^c \mapsto z^{py_{U}} U^c,  z^{py_{D}} D^c$. Thus we require that $y_Q+ y_{D} =
 -y_Q - y_{U} = \pm \frac{q}{p}$. Locally, these are precisely the same relations that we
 require in the SM, up to an unphysical overall
 rescaling.

Thus, locally, all such models are
equivalent to models with $p=1$. But models with $p=1$ (with the Higgs
boson treated as an additional, $CP$-even, singlet, scalar matter field with
arbitrary couplings \cite{Contino:2010mh}) contain the SM
(or rather a low-energy limit thereof) as a special case and thus
can be consistent with data. So for any $p$, at the local level, there exists a
choice of parameters in the corresponding NLSM that is consistent with
particle physics data. 
\subsection{Custodial symmetries \label{sec:sm}}
Though local considerations suggest that a NLSM based on $G/H_{p,q}$ can fit the data for any $p$, it is
clearly far less satisfactory than the SM as regards its
predictivity. Most seriously, while the Higgs boson is an integral
part of the SM, in an NLSM description we are forced to 
arbitrarily
include an additional scalar matter field whose couplings are tuned to
be close to those of the SM Higgs boson. Even if we are willing to overlook these issues regarding the Higgs,
there are two more successful predictions of the SM which, though they can be
accommodated by any model, are not predicted in a generic
NLSM. The first of these is the $W-Z$ boson mass ratio, which is
fixed in the SM, but is arbitrary in the $G/H_{p,q}$ NLSM. Indeed, the
gauge boson masses are determined by specifying a $G$-invariant metric
on $G/H_{p,q}$. In the SM, the metric is fixed to be the round metric on $S^3$,
which is unique up to an overall normalization, so that the $W-Z$ mass
ratio is fixed. But in a generic NLSM, we may pick an arbitrary
$G$-invariant metric on $G/H_{p,q}$. Such metrics may be classified as
follows \cite{KobN}:
as $G$ acts almost effectively on
$G/H_{p,q}$ (i.e.\ the subgroup of $G$ that fixes all elements of
$G/H_{p,q}$ is discrete), the $G$-invariant metrics on $G/H_{p,q}$ are in 1-1
correspondence with those inner products on $\mathfrak{g}/\mathfrak{h}_{p,q}$ which are invariant for the adjoint action of $H_{p,q}$. 
The adjoint representation of $SU(2)\times U(1)$ restricts to $H_{p,q}$ as
$3 \oplus 1 \rightarrow 2q \oplus -2q \oplus 0 \oplus 0$, where we denote
representations on the LHS by their dimension and on the RHS by
their $H_{p,q}$ charge. There are thus 2 such independent inner
products, each with an arbitrary overall normalization,
leading to an arbitrary $W-Z$ mass ratio.

As is well-known, the particular mass ratio that obtains in the SM can
be understood via custodial symmetry \cite{Sikivie:1980hm}: the vacuum manifold
$S^3$ is invariant under a larger action of $SU(2)\times SU(2)$ (with
the original $U(1)$ included in the second $SU(2)$), broken to the diagonal $SU(2)$ by the Higgs
VEV. Since the adjoint representation of $SU(2)\times SU(2)$ restricts as $(3,1)
\oplus (1,3) \rightarrow 3 \oplus 3$, there is just one invariant
inner product, up to an overall normalization. 

We claim that a similar construction can be
applied to the $G/H_{p,q}$ NLSM only in the cases $p\in \{1,2\}$. 
Indeed, to do so requires us to find a metric on $G/H_{p,q}$ which is
invariant not just under $SU(2)\times U(1)$, but rather under the
larger $SU(2)\times SU(2)$. Now, this larger group need not act
effectively on $G/H_{p,q}$, but the
subgroup $N = \{g \in SU(2)\times SU(2):
g.x = x \; \forall \; x \in G/H_{p,q}\}$, being the kernel of a homomorphism from
$SU(2)\times SU(2)$ to $\mathrm{Sym} (G/H_{p,q})$, will necessarily be normal, and furthermore
$(SU(2)\times SU(2))/N$ will act effectively on $G/H_{p,q}$. Now, the normal subgroups
of $SU(2)$ are $\{\{e\},\mathbb{Z}/2,SU(2)\}$ and projecting $N$ to either $SU(2)$-factor must give one of these. But neither projection can be $SU(2)$, as otherwise $(SU(2) \times SU(2))/N$ would have dimension at most 3, which is incompatible with the known restriction of the action to the subgroup
$SU(2)\times U(1)$ (in which both factors have a non-trivial action),
so in fact $N$ must be a subgroup of the centre $\mathbb{Z}/2 \times \mathbb{Z}/2$. Thus the $SU(2)\times SU(2)$-action is almost effective and the isometry group of the desired
metric has dimension at least 6. 
But it is a theorem \cite{Kob} that the isometry group of a
compact Riemannian $3$-manifold has dimension at most 6, equalling 6 only
if it is $S^3$ or $\mathbb{R}P^3$, corresponding to $p=1$ or $p=2$ respectively.

Thus, we see that a custodial symmetry protecting $W$ and $Z$ masses
may also be imposed in the topologically
non-trivial case $p=2$. Here $G/H_{2,1} \cong \mathbb{R}P^3 \cong SO(3)$ has an action of $SO(3) \times SO(3)$ by $(g,h).x = g x h^{-1}$,
with the $SU(2)\times SU(2)$-action being the one
induced by
the covering map from $SU(2) \rightarrow SO(3)$; the LSM model
(respectively composite Higgs model) in
\S~\ref{sec:lsm} (respectively \S~\ref{sec:ch}) give explicit realisations.

The second successful prediction of the SM involves the couplings of
gauge bosons to fermions.  Again, for a generic $G/H_{p,q}$ NLSM,
deviations in these couplings may occur. 
In a concrete theory of flavour such as partial compositeness (which
has the {\em desiderata} of explaining much of the hierarchical structure of Yukawa couplings whilst suppressing potential
flavour-changing effects) \cite{Kaplan:1991dc} this is not so much of a problem, because
the SM fermions are largely elementary, being weakly mixed with the
strong sector. Deviations are therefore expected to be small. The biggest problem
arises in the coupling of the $Z$ boson to left-handed
bottom quarks, since the latter belongs to the same $SU(2)$ doublet as the
top quark, and is forced to be sizably mixed with the sigma model sector
in order to accommodate the large top quark mass. But this too can be
protected by a symmetry in the $p=1$ case \cite{Agashe:2006at}. 
The required group is 
$(SU(2) \times SU(2)) \rtimes
\mathbb{Z}/2$, where $\mathbb{Z}/2$
permutes the two $SU(2)$s \cite{Gripaios:2014pqa}. A similar protection can also be
achieved when $p=2$, by using the action of $(SO(3)
\times SO(3)) \rtimes \mathbb{Z}/2$ on $G/H_{2,1} \cong \mathbb{R}P^3
\cong SO(3)$, where the $\mathbb{Z}/2$ acts by inversion on the group
$SO(3)$.
\subsection{Global properties \label{sec:glo}}
Problems with consistency with the data arise when we consider the
global properties of a $G/H_{p,q}$ NLSM. The first obvious question is
whether the presence of topologically-stable string solutions is
consistent with astrophysical data. In fact, this issue is hard to settle.
One would expect a network of
strings to form in the early Universe during the cosmological EW phase
transition \cite{Vilenkin:2000jqa}, but the purely gravitational
effects of such strings are of order $v^2/M_P^2 \sim 10^{-34}$ and are
utterly negligible. However, such a string necessarily features quark and lepton zero
modes localised on its core, which lead to the formation of superconducting currents (as
well as baryon- and lepton-number violation) in
the presence of astrophysical magnetic fields \cite{Witten:1984eb}. A
number of resulting astrophysical signatures have been discussed
(in, among others, the microwave background, radio bursts, cosmic rays, and
galactic and stellar dynamics; for a review, see \cite{Hindmarsh:1994re}) but
there seems to be no consensus that they lead to robust constraints
on EW-scale strings. 

A much more serious (and easier to settle) problem occurs when we
  consider fermion masses. We saw above that, for quarks say, these
  require us to satisfy the conditions $y_Q+ y_{D} =
 -y_Q - y_{U} = \pm \frac{q}{p}$, which, locally, are no different
 from the
 conditions one obtains in the SM. But there is a problem globally, which is that the hypercharges must be
 integers, in order that the fermions carry {\em bona fide}
 representations of $SU(2) \times U(1)$. This, together with the
 requirement that $p,q$ be coprime, implies that one cannot write
$H_{p,q}$-invariant fermion mass terms unless $p=1$. Hence, one
obtains a gross conflict with data for $p \neq 1$.

There is a simple way to avoid this unfortunate conclusion, which is
to  change the
representation content of the fermion fields. In particular, mass
terms are allowed if the fermions are
assigned to higher-dimensional representations of $SU(2) \times
U(1)$. But this has the undesirable knock-on effect of either
requiring additional unobserved fermion states (to fill out the
higher-dimensional multiplets), or of modifying
the observed couplings of existing fermions to gauge bosons. Neither
is phenomenologically viable.
\section{Linear sigma models \label{sec:lsm}}
Even though we have seen that models with topologically non-trivial vacuum
  manifold are incompatible with data, it is interesting to see how
  they could have arisen from physically-sensible, UV-complete
  models. As a first, example, we consider linear sigma models.
A model in which the Higgs
field of the SM is replaced by a scalar carrying the $(2j+1,q)$
representation of $SU(2)\times U(1)$ (where the spin $j$
representation of $SU(2)$ is labelled by its dimension $2j+1$) leads
(for a suitable choice of orbit for the VEV) to a vacuum manifold
homeomorphic to $L(p,1)$, with $p = 2j/\mathrm{gcd} (2j,q)$. Either the
arguments given in \S \ref{sec:nlsm} or an explicit calculation shows that
the pattern of couplings of gauge bosons to themselves and to fermions
is exactly as in the SM,
 but, since
custodial symmetry is not respected, there is a gross violation of the
$W-Z$ mass ratio for $2j\neq 1$, with $\frac{m_W^2}{m_Z^2} =
\frac{g_2^2}{2j(g_2^2 + g_1^2)}$ at tree-level, where $g_2$ (respectively
$g_1$) denotes the SM value of the $SU(2)$ (resp. $U(1)$) gauge coupling.

For a model yielding the correct tree-level value of $W-Z$ mass ratio,
we replace the SM Higgs field by a
real scalar, $\Phi$, carrying a bi-triplet, $(3,3)$, representation of $O(3)\times
O(3)$. We employ a matrix notation where the action of 
$(L,R) \in O(3)\times
O(3)$ is given by $\Phi \mapsto L \Phi R^T$. We may then
define an action of $SU(2)\times SU(2)$ on $\Phi$ via the
usual covering map $\pi: SU(2) \rightarrow SO(3) < O(3)$ and so there is
also an action of the SM electroweak gauge group $SU(2) \times
U(1)$ (with $U(1)$ a subgroup of the other $SU(2)$). 

The most general, renormalizable, $O(3)\times O(3)$-invariant
potential for $\Phi$ is a linear combination of $\mathrm{tr} \Phi^T
\Phi, \mathrm{tr} (\Phi^T
\Phi)^2,$ and $(\mathrm{tr} \Phi^T
\Phi)^2,$ 
which may be written in the form
$V(\Phi) = a \mathrm{tr} (\Phi^T
\Phi - c^2)^2 + b (\mathrm{tr} \Phi^T
\Phi - 3c^2)^2$,
where $a,b,c^2 \in \mathbb{R}$ and $a,b>0$. Written thus, it is clear that the minimum of the
potential lies at $\Phi^T \Phi = c^2 1$ (for $c^2 >0$).
 Not only does this tell us that the unbroken subgroup of $O(3)\times O(3)$ is indeed the diagonal $O(3)$ (since we can
choose $\langle \Phi \rangle = c 1$ such that $\langle \Phi \rangle
\mapsto c L R^T$ which equals $\langle \Phi \rangle = c 1$ iff. $L =
R$), but it also tells us that the vacuum manifold is homeomorphic to $O(3)$, whose component connected to the identity is homeomorphic to $SO(3) \cong \mathbb{R}P^3$.
Since $(3,3)$ restricts to $1\oplus 3 \oplus 5$ under $O(3) <
O(3)\times O(3)$, we expect to find a spectrum in the Higgs sector
consisting of the 3 massless Goldstone bosons, together with a further 6
massive states, of which at least 5 are degenerate in mass. Indeed, if
we expand $V(\Phi)$ to quadratic order using $\Phi = (c+h)1 + S + A$,
where $S$ is traceless and symmetric and $A$ is antisymmetric, we
find that the first term in the potential generates
equal masses for $S$ and $h$, while the second term (because it is, in
fact, $O(9)$ symmetric) generates a mass
for $h$ only. 

Consistently with our arguments in \S \ref{sec:glo}, one sees
  that there is
no way to generate acceptable masses for SM fermions in this model via
Yukawa couplings, since $\Phi$ can only couple to
pairs of $SU(2)_L$ doublets or pairs of singlets, rather than to a
doublet and a singlet as in the SM. If one reinstates the SM Higgs to fix this, as in the model of
\cite{Georgi:1985nv}, one may show that the vacuum manifold is restored
to $S^3$.
\section{A composite Higgs model \label{sec:ch}}
It is also possible to achieve a topologically non-trivial vacuum
manifold in models in which the hierarchy problem is solved by making
the Higgs boson a composite of some new TeV-scale dynamics. Indeed, in the
most favoured among such models \cite{Agashe:2006at}, the Higgs is a
pseudo-Goldstone boson,
taking values in a homogeneous space
$SO(5)/O(4) \simeq \mathbb{R}P^4$. Due to the presence of couplings to
gauge bosons and
fermions, the dynamics cannot be fully $SO(5)$ invariant and
so the low-energy effective lagrangian contains a potential for the
Higgs field. This is a real-valued, $SU(2)\times U(1)$-invariant function on $\mathbb{R}P^4$, with
the vacuum manifold being given by the level set of the minimum. The
specific form of the function is determined by the uncalculable strong
dynamics, but let us suppose, for the sake of illustration, that it takes the form
$V = \frac{x_1^2}{x_1^2 + x_2^2 + x_3^2 + x_4^2 +
  x_5^2}$, where $x_i, i\in \{1,2,3,4,5\}$ are coordinates in
$\mathbb{R}^5$. This is well-defined on $\mathbb{R}P^4$, the space of lines through the
origin in $\mathbb{R}^5$, and is, moreover, smooth and invariant under
the larger group $O(4)$. We have that $0\leq V \leq 1$, with the
maximum at the point $x_2=x_3=x_4=0$ and the minimum at $x_1 = 0$,
corresponding to the level set $\mathbb{R}P^3$. 

In accordance with our general result, choosing such a minimum as
  vacuum must imply
vanishing fermion masses; this can be confirmed by comparing with, {\em e.g.}, formul\ae\ for the top quark
mass in explicit models in \cite{DeSimone:2012fs}.
\section{Groups locally isomorphic to $SU(2)\times U(1)$ \label{sec:u2}}
Finally, we consider gauge groups that are locally, but not globally,
isomorphic to $SU(2)\times U(1)$. Such groups need not be connected, in
which case the possibilities are infinite, but all feature domain-wall
type solutions that are potentially dangerous from a cosmological
point of view. If we restrict our
attention to the component connected to the
identity, then the possibilities are finite in number, given by quotients of the
universal cover $SU(2) \times \mathbb{R}$ by a discrete subgroup of the centre
$\mathbb{Z}/2 \times \mathbb{R}$. Of the five possibilities, the only
ones other than $SU(2) \times U(1)$ admitting doublet irreducible representations (as carried by quarks and
leptons) are $SU(2) \times \mathbb{R}$ and $U(2)$. The former has
subgroups isomorphic to $\mathbb{R}$ and is disfavoured by the
apparent quantization of hypercharge; similar arguments to those
given in \S \ref{sec:21} show that the vacuum manifold is always
homeomorphic to $S^3$ in this case. The latter has $U(1)$ subgroups
given by $H_{r,s} = \{\mathrm{diag} (e^{ir\theta},e^{is\theta})\}$,
with $r,s \in \mathbb{Z}$ and coprime, is homeomorphic to
$L(r+s,1)$, and also admits topologically-stable strings.
\section*{Acknowledgements}
BG acknowledges
the support of STFC grant ST/L000385/1 and both authors acknowledge
the support of King's College,
Cambridge. 
%
\end{document}